\documentclass{article}
\usepackage{arxiv}
\usepackage{graphicx}
\usepackage[dvipsnames]{xcolor}
\usepackage{hyperref}

\usepackage{bm}

\usepackage{listings, xcolor}

\definecolor{verylightgray}{rgb}{.97,.97,.97}

\lstdefinelanguage{Solidity}{
  keywords=[1]{anonymous, assembly, assert, balance, break, call, callcode, case, catch, class, constant, continue, contract, debugger, default, delegatecall, delete, do, else, emit, event, export, external, false, finally, for, function, gas, if, implements, import, in, indexed, instanceof, interface, internal, is, length, library, log0, log1, log2, log3, log4, memory, modifier, new, payable, pragma, private, protected, public, pure, push, require, return, returns, revert, selfdestruct, send, storage, struct, suicide, super, switch, then, this, throw, transfer, true, try, typeof, using, value, view, while, with, addmod, ecrecover, keccak256, mulmod, ripemd160, sha256, sha3, constructor}, 
  keywordstyle=[1]\color{blue}\bfseries,
  keywords=[2]{address, bool, byte, bytes, bytes1, bytes2, bytes3, bytes4, bytes5, bytes6, bytes7, bytes8, bytes9, bytes10, bytes11, bytes12, bytes13, bytes14, bytes15, bytes16, bytes17, bytes18, bytes19, bytes20, bytes21, bytes22, bytes23, bytes24, bytes25, bytes26, bytes27, bytes28, bytes29, bytes30, bytes31, bytes32, enum, int, int8, int16, int24, int32, int40, int48, int56, int64, int72, int80, int88, int96, int104, int112, int120, int128, int136, int144, int152, int160, int168, int176, int184, int192, int200, int208, int216, int224, int232, int240, int248, int256, mapping, string, uint, uint8, uint16, uint24, uint32, uint40, uint48, uint56, uint64, uint72, uint80, uint88, uint96, uint104, uint112, uint120, uint128, uint136, uint144, uint152, uint160, uint168, uint176, uint184, uint192, uint200, uint208, uint216, uint224, uint232, uint240, uint248, uint256, var, func, void, ether, finney, szabo, wei, days, hours, minutes, seconds, weeks, years}, 
  keywordstyle=[2]\color{teal}\bfseries,
  keywords=[3]{block, blockhash, coinbase, difficulty, gaslimit, number, timestamp, msg, data, gas, sender, sig, value, now, tx, gasprice, origin}, 
  keywordstyle=[3]\color{violet}\bfseries,
  keywords=[4]{@check, @invariant, @never, @set_restricted, var=}, 
  keywordstyle=[4]\color{MidnightBlue}\bfseries,
  identifierstyle=\color{black},
  sensitive=false,
  comment=[l]{//},
  morecomment=[s]{/*}{*/},
  commentstyle=\color{gray}\ttfamily,
  stringstyle=\color{Thistle}\ttfamily,
  morestring=[b]',
  morestring=[b]"
}

\usepackage{amsmath}
\usepackage{tikz}
\usetikzlibrary{arrows, automata, backgrounds, calendar, positioning, calc, shapes, snakes, fit}
\usepackage{amssymb}
\usepackage{color}
\usepackage{adjustbox}
\definecolor{compadd}{rgb}{.0,.5,.3}
\definecolor{amber}{rgb}{1.0, 0.75, 0.0}
\definecolor{darkorange}{HTML}{BB4400}
\definecolor{darkgreen}{HTML}{006400}
\usepackage{hyperref}
\makeatletter
\pgfdeclareshape{document}{
\inheritsavedanchors[from=rectangle] 
\inheritanchorborder[from=rectangle]
\inheritanchor[from=rectangle]{center}
\inheritanchor[from=rectangle]{north}
\inheritanchor[from=rectangle]{south}
\inheritanchor[from=rectangle]{west}
\inheritanchor[from=rectangle]{east}
\backgroundpath{
\southwest \pgf@xa=\pgf@x \pgf@ya=\pgf@y
\northeast \pgf@xb=\pgf@x \pgf@yb=\pgf@y
\pgf@xc=\pgf@xb \advance\pgf@xc by-10pt 
\pgf@yc=\pgf@yb \advance\pgf@yc by-10pt
\pgfpathmoveto{\pgfpoint{\pgf@xa}{\pgf@ya}}
\pgfpathlineto{\pgfpoint{\pgf@xa}{\pgf@yb}}
\pgfpathlineto{\pgfpoint{\pgf@xc}{\pgf@yb}}
\pgfpathlineto{\pgfpoint{\pgf@xb}{\pgf@yc}}
\pgfpathlineto{\pgfpoint{\pgf@xb}{\pgf@ya}}
\pgfpathclose
\pgfpathmoveto{\pgfpoint{\pgf@xc}{\pgf@yb}}
\pgfpathlineto{\pgfpoint{\pgf@xc}{\pgf@yc}}
\pgfpathlineto{\pgfpoint{\pgf@xb}{\pgf@yc}}
\pgfpathlineto{\pgfpoint{\pgf@xc}{\pgf@yc}}
}
}
\makeatother
\tikzstyle{docc}=[%
   draw,
   thick,
   align=center,
   color=black,
   shape=document,
   minimum width=10mm,
   minimum height=14.1mm,
   shape=document,
   inner sep=4ex,
]

\newsavebox{\mybox}
\newcommand{\roundbox}[1]{%
  \tikz[baseline=-1ex]%
  \node[%
  inner sep=1.5pt,
  draw=black,
  fill=black,
  text=white,
rounded corners=2.5pt]{#1};
}

\newcommand{\coderef}[1]{%
  \begingroup%
  \scriptsize\ttfamily%
  \roundbox{\autoref{#1}}%
  \endgroup%
}

\newcount\mymark
\makeatletter
\def\codelabel#1{%
  \global\advance\mymark by 1%
  \protected@write \@auxout {}{\string \newlabel {#1}{{\the\mymark}{}}}%
  \makebox[17pt][r]{{\scriptsize\roundbox{\the\mymark}~}}%
}
\makeatother

\lstdefinestyle{base}{
  basicstyle=\ttfamily\color{gray},
  moredelim=[is][\underbar]{@}{@},
}

\def\postbreak{\raisebox{0ex}[0ex][0ex]{\ensuremath{\hookrightarrow\space}}}
\begin{document}
\title{Annotary: A Concolic Execution System for Developing Secure Smart Contracts}
%
%

\author{Konrad Weiss\\
Fraunhofer AISEC\\
\texttt{konrad.weiss@aisec.fraunhofer.de}
\And
Julian Sch\"{u}tte\\
Fraunhofer AISEC\\
\texttt{julian.schuette@aisec.fraunhofer.de}
}

%


\maketitle              

\begin{abstract}
Ethereum smart contracts are executable programs, deployed on a peer-to-peer network and executed in a consensus-based fashion. Their bytecode is public, immutable and once deployed to the blockchain, cannot be patched anymore. As smart contracts may hold Ether worth of several million dollars, they are attractive targets for attackers and indeed some contracts have successfully been exploited in the recent past, resulting in tremendous financial losses. The correctness of smart contracts is thus of utmost importance. While first approaches on formal verification exist, they demand users to be well-versed in formal methods which are alien to many developers and are only able to analyze individual contracts, without considering their execution environment, i.e., calls to external contracts, sequences of transaction, and values from the actual blockchain storage. In this paper, we present Annotary, a concolic execution framework to analyze smart contracts for vulnerabilities, supported by annotations which developers write directly in the Solidity source code. In contrast to existing work, Annotary supports analysis of inter-transactional, inter-contract control flows and combines symbolic execution of EVM bytecode with a resolution of concrete values from the public Ethereum blockchain. While the analysis of Annotary tends to weight precision higher than soundness, we analyze inter-transactional call chains to eliminate false positives from unreachable states that traditional symbolic execution would not be able to handle. We present the annotation and analysis concepts of Annotary, explain its implementation on top of the Laser symbolic virtual machine, and demonstrate its usage as a plugin for the Sublime Text editor.
\end{abstract}

\section{Introduction}
Smart contracts are small programs, executed by all verifying nodes of a blockchain as part of a consensus protocol. The idea of smart contracts is to distribute not only data but also computation to a set of potentially untrusted peers, in order to create distributed applications (''DApps'') that are not governed by a single party and operate correctly and reliably, as long as the majority of the blockchain network sticks to the protocol. In that sense, smart contracts implement the core business logic of DApps and are responsible for moving digital currency from one account (i.e., user) to another.

Ethereum, the most popular public implementation of the concept of smart contracts, is a permissionless public blockchain that uses the above concepts to create a digital currency called Ether, as well as a general-purpose distributed computing engine with a quasi-Turing complete execution model.~\cite{buterin2013ethereum}. Ether is publicly tradable similar to Bitcoin but also serves as the payment method for code execution of smart contracts in the Ethereum Virtual Machine (\textbf{EVM}), typically written in the programming language Solidity~\cite{solidity} and compiled into the EVM bytecode format which is then deployed to the peer-to-peer network.

In some applications, the amount of Ether controlled by a smart contract is enormous. 
From a security perspective, smart contract code can thus be regarded similar to code of smart card applets: the code implements simple functionality in a well-defined and constrained environment and is thus easy to verify, but errors in that code are not tolerable as extremely high values are at stake. At the same time, once deployed to the public, it is almost impossible to roll out security patches.
Blockchains and DApps are created in a rapidly evolving industry where time-to-market is crucial, and smart contract developers rarely have a background in writing highly critical code or experience with formal verification methods. Various severe incidents have happened in the past, where vulnerabilities in smart contracts allowed to lock in or withdraw significant amounts of Ether from popular DApps. To name only a few, this includes the PoWHCoin bug, the first Parity bug (153,037 ETH stolen)~\cite{paritymultisig1}, the second Parity bug (513,774.16 ETH frozen)~\cite{paritymultisig2}, and the DAO hack (3.6 mio. ETH stolen)~\cite{theDao} which finally lead to a hard fork of the Ethereum blockchain. These incidents suggest that writing secure smart contracts is challenging and effectively supporting developers in avoiding vulnerabilities is a necessity. Rigid formal verification methods have been proposed in the past \cite{whytosolc} but later dismissed, as they put too high demands on developers who are no experts in this field. Simple static analysis approaches, on the other hand, help to avoid simple programming errors but are far from being precise enough to discover subtle flaws -- especially those manifesting in the interaction between multiple contracts.

In this paper, we introduce \emph{Annotary}, a concolic execution tool that supports Solidity developers in writing error-free smart contracts. In contrast to other tools which focus on searching predefined vulnerability patterns, we take a developer-centric perspective and allow developers to express their expectations in the form of annotations directly in the Solidity code. \emph{Annotary} then conducts a concolic execution analysis of the compiled EVM bytecode against these annotations and informs the developer about potential violations -- currently in the form of a plugin for the Sublime editor. We advance the state of the art in EVM analysis by including interactions between contracts and along chains of transactions in the analysis and make the following contributions

\begin{enumerate}
    \item extend concolic analysis of EVM bytecode to properly span contract interactions and sequences of transactions.
    \item a backward-compatible extension of the Solidity language by annotations which allows developers to state verifiable properties
    \item a proof-of-concept implementation of \emph{Annotary}, including a Sublime Text plugin
\end{enumerate}

\section{Background}

Although at a syntactical level, Solidity resembles C or JavaScript, its execution model has some peculiarities that require further discussion. Furthermore, we will provide some background on Mythril, a vulnerability scanning tool for Ethereum smart contracts, that we significantly extended in the process of developing \emph{Annotary}.

\subsection{Solidity and Smart Contracts}
Solidity is a high-level language for implementing smart contracts and targets the Ethereum Virtual Machine (EVM) platform. It is statically typed, supports multiple inheritance, libraries, complex user-defined types, contracts as members, overloading and overwriting, abstraction and interfaces, as well as encapsulation through visibility modifiers. Solidity's contract-orientation appears similar to object-oriented languages, using the \texttt{contract} keyword instead of \texttt{class}. However, in contrast, to truly object-oriented languages, such type definitions do not end up in the actual bytecode which consequently only includes instantiations of contracts and their respective functions~\cite{solidity}. A special contract-creation transaction is used to invoke the ''constructor'' and as a result, the contract is instantiated and assigned a public address which only holds the code that can be called by transactions. It is also important to note that contracts created from the same code basis do not share any data or (static) functions.

\autoref{lst:smallcontract} illustrates some typical concepts of the Solidity language, including inheritance and two different ways to declare constructors. The constructor in ~\coderef{code:l1} is declared by naming the function equal to its contract (analog to languages like Java), while~\coderef{code:l3} uses the newer \texttt{constructor} keyword (analog to JavaScript, albeit Solidity merely treats \texttt{constructor} as a function modifier), which became mandatory in version 5.0~\cite{solidity5} of Solidity to avoid vulnerabilities related to simply misspelling function names. The example also shows two patterns which are common in smart contracts: first, the constructor keeps track of the owner who originally deployed the contract by assigning the associated 20-byte address~\coderef{code:l2} passed to the constructor to the \texttt{owner} field. This allows the contract to later distinguish between calls that are made by its original owner or by anyone else.
Second, the contracts defines a nameless \emph{default function}~\coderef{code:l4} that is called when callers invoke the contract without referring to a specific function. In this case, the \texttt{require}-statement will roll back the transaction if the sender attempts to send any Ether (\texttt{msg.value}) to a non-existing function. To address a specific function in a transaction, it must include the function identifier, which is computed as the first four most significant bytes of the keccak256 hash of the function signature.\newline
\vspace*{.5cm}
\begin{lstlisting}[language=Solidity,caption={Solidity smart contract example},captionpos=b, ,label={lst:smallcontract}]
contract A {
  address owner;
  function A(){ owner = msg.sender;(*\codelabel{code:l2}*)}(*\codelabel{code:l1}*)
}
contract B is A{
  uint variable;
  function constructor(){ variable = 1;}(*\codelabel{code:l3}*)
  function setVar(uint var1){ ... }
  function() payable { require(msg.value == 0);}(*\codelabel{code:l4}*)
}
\end{lstlisting}
\paragraph{\textbf{Entities and Interactions}}
Ethereum is a distributed system building a singleton computer with accounts as entities and transactions referring to accounts as the smallest units of computation. Accounts are identified by a 160-bit address, have a balance of Ether, a transaction counter, and two possibly empty fields: the associated bytecode and storage state. \textbf{Wallets} are contracts with empty bytecode and their 160-bit address is the hash of their public key. The holder of the private key signs transactions proving its origin to be the Wallet. Transactions to these accounts can only transfer Ether. Accounts with associated bytecode are \textbf{contract accounts} and receive their address in a deterministic process when a contract creation transaction is sent to the network. The construction-bytecode is run, and the resulting state of the contract is some possibly non-empty storage state and the runtime-bytecode.\newline
\textbf{Transactions} are signed data packets that represent a message to an account by specifying its address in the \texttt{to}-field or a contract-creation transaction if the content is 0. \textbf{Messages} can be the result of a transaction or of subsequent deterministic calls between contracts when bytecode is executed. They are unsigned blocks of data sent from one account to another still associated with the verified initial sender of the transaction. If an account with non-empty bytecode receives a message, an instance of the EVM is started with the target account's bytecode and the message data as input. Returned data is passed to the calling EVM context or returned as transaction result.

\subsection{EVM and Bytecode}\label{subsec:backevmbytecode}
EVM has a simple stack-based architecture with a word size of 256 bit that allows to directly map keccak-256 hashes to addresses. A predefined finite resource called \texttt{gas} must be assigned to each transaction and serves as the unit for computational effort that is consumed by each EVM instruction. It thus helps to prevent Denial-of-Service (DoS) attacks by stopping the execution when it is depleted, making the EVM a \textit{quasi}-Turing-complete machine that has a Turing-complete instruction set but can only execute a limited number of statements~\cite{yellowpaper}.
The simplicity of the EVM bytecode with its 70 main instructions made the EVM a popular target for formal verification projects~\cite{kevm,Park2018,Bhargavan2016}.
During execution the EVM maintains three main types of memory that are also relevant for the analysis by \emph{Annotary}:

The \textbf{world state} ${\bm{\sigma}}$ is a mapping from 160-bit addresses ${\bm{a}}$ to account states and is kept in a Merkle Patricia tree that represents the result of executing all transactions saved on the Ethereum blockchain. A mapped account state ${\bm{\sigma[a]}}$ contains the \textbf{balance} ${\bm{\sigma[a]_b}}$ in Wei, the smallest sub-unit of Ether ($10^{18}$ Wei = 1 Ether), the \textbf{storage} ${\bm{\sigma[a]_s}}$ as mapping from 256-bit integer values to 256-bit integer values ${\bm{\sigma[a]_s}:2^{256}\rightarrow2^{256}}$, and the immutable \textbf{runtime bytecode} ${\bm{\sigma[a]_c}}$ of an account that is executed in the case of message receipt.

The \textbf{execution environment} ${\bm{I}}$ contains data that is fixed during message processing, including the address of the current message recipient ${\bm{I_a}}$ whose code is executed, and the sender of the message ${\bm{I_s}}$. ${\bm{I_o}}$ is the account associated with the original transaction and may differ from ${\bm{I_s}}$ for inter-contract messages. ${\bm{I_d}}$ contains the input data for the current execution, such as function parameters. ${\bm{I_v}}$ contains the value of Ether in Wei transfered from the sender ${\bm{I_s}}$ to the recipient ${\bm{I_a}}$. ${\bm{I_b}}$ contains the runtime bytecode of ${\bm{I_a}}$ that is executed and ${\bm{I_H}}$ stores the header of the block that the current transaction will be mined in.

The \textbf{machine state} ${\bm{\mu}}$ contains the variable and volatile part of the computation held only during message processing. These include the volatile operand LIFO stack ${\bm{\mu_s}}$ with 256-bit words and a byte-addressable heap memory ${\bm{\mu_m}}$ used for more complicated computation or larger chunks of data. It further holds the program counter ${\bm{\mu_{pc}}}$ and the output byte-array ${\bm{\mu_o}}$ of the execution. Changes of the execution are not persisted if not send or returned to a different account or stored in ${\bm{\sigma[a]}}$.

\subsection{Mythril and the \textsc{Laser}-SVM}
Mythril~\cite{mythril} is an open-source security analysis tool for Ethereum smart contracts and serves as the foundation for \emph{Annotary}\footnote{As Mythril is under active development, this paper refers to the commit hash \url{github.com/ConsenSys/mythril-classic/commit/b5afa9ff1aa2b5dc8863d29aa9e0a24b34eb4747} of the project}\newline
It uses \textsc{Laser}-SVM, an internal symbolic virtual machine, to explore smart contract bytecode in a depth-first search fashion over the control flow graph (CFG). For this, it operates on a representation of ${\bm{\sigma}}$, ${\bm{I}}$ and ${\bm{\mu}}$. Values that are unknown during execution are represented as symbolic variables, and the explored execution paths are transformed into \emph{path conditions}, i.e., constraint systems over the symbolic variables along the respective execution path. Mythril runs vulnerability detection modules that inspect the explored executions states for known vulnerability patterns and attempts to compute concrete input values leading to the execution of the vulnerability by solving the respective constraint system using the Z3 SMT solver~\cite{z3}. As the EVM uses 256-bit operands for computation, Mythril uses a bit-vector algebra at a fixed size of 256 bits to model arithmetic operations and boolean algebra. While Mythril and especially its \textsc{Laser}-SVM provide a good basis of concolic/symbolic execution for EVM bytecode, Mythril's goal is not to allow analysis of specifiable properties.
Mythril lacks the following capabilities which \emph{Annotary} aims to provide:
\begin{itemize}
  \item The ability to let developers specify invariants and assertions in the contract and the ability to verify them \emph{before} the potentially vulnerable contract is irrevocably deployed.
  \item A model of EVM instructions and execution semantics for inter-contract- and inter-transactional control flows.
  \item Reachability analysis of transaction sequences to reduce false positives.
  \item Symbolic execution of contract constructors with parameters.
\end{itemize}

\section{Annotation Driven Concolic Analysis}
Rather than exploitation, \emph{Annotary} aims at secure development while expanding the analysis scope to inter-contract and inter-transactional analyses. We begin this section by outlining the overall system and then detailing the main aspects of the analysis.
\tikzstyle{doc}=[minimum height=4em,minimum width=3em,draw, fill=white]
\tikzset{ documents/.pic = {
      \node[doc] (-behind) {};
      \node[doc, below left = 4pt and 4pt of -behind.north east] (doc2) {};
      \node[doc,  below left = 4pt and 4pt of doc2.north east] (-front) {\rotatebox{90}{#1}};
    }}
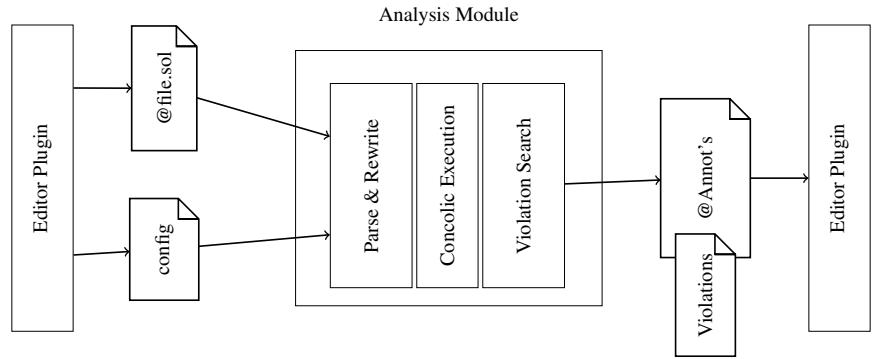
\begin{figure}[h!]
    \centering
    \resizebox{.7\textwidth}{!}{
    \begin{tikzpicture}

    \node[rectangle,draw, minimum height=15em,minimum width=3em] (an) {\rotatebox{90}{Editor Plugin}};

    \node[docc, right=of an, yshift=4.4em,inner sep=1.2em] (A) {\rotatebox{90}{\textbf{@}file.sol}};
    \node[docc, below=of A,yshift=.5em,inner sep=1.2em] (B) {\rotatebox{90}{config}};

    \node[rectangle,draw, right=of an, xshift= 8em, minimum height=12.5em,minimum width=15em,] (myth) {};

    \node[above= of myth, yshift=-2em] () {Analysis Module};

    \node[rectangle,draw, above=of myth, minimum height=10em,minimum width=4em, yshift=-14.5em, xshift=-3.8em] (il) {\rotatebox{90}{Parse \& Rewrite}};
    \node[rectangle,draw, right=of il, minimum height=10em,minimum width=3em, xshift=-2.655em] (laser) {\rotatebox{90}{Concolic Execution}};
    \node[rectangle,draw, right=of laser, minimum height=10em,minimum width=4em, xshift=-2.655em] (ol) {\rotatebox{90}{Violation Search}};

    \node[docc, right=of myth] (a) {\rotatebox{90}{\textbf{@}Annot's}};

    \node[docc,below=of a, fill=white!40, inner sep=2ex, yshift=4em] (v) {\rotatebox{90}{Violations}};

    \node[rectangle,draw, minimum height=15em,minimum width=3em,right=of a] (anv) {\rotatebox{90}{Editor Plugin}};

    \draw[->,thick] (an.71) -- (A);
    \draw[->,thick] (an.-68) -- (B);
    \draw[->,thick] (A) -- (il.130);
    \draw[->,thick] (B) -- (il.-130);

    \draw[->,thick] (ol) -- (a);
    \draw[->,thick] (a) -- (anv);

    \end{tikzpicture}

    }
    \caption{\emph{Annotary's} architecture with Solidity files undergoing analysis and violations reported to the editor plugin.}\label{fig:pluginattached}
\end{figure}
Figure~\ref{fig:pluginattached} shows how \emph{Annotary's} editor plugin passes source and configuration files to the analysis component and receives found annotation violations for visualization.\newline

\subsection{Annotations}
\emph{Annotary} specifies a set of annotations which developers can use to express invariants and restrictions directly in the Solidity source code. These annotations will then be translated into constraints or injected as asserts and analyzed to become part of the constraint system for an execution path. As annotations may include expressions, as well as references to Solidity functions and members, they require a separate compilation pass in addition to compilation of the actual source code. This is done by the annotation processor which takes annotations as input and translates them into EVM instructions by rewriting the original contract code. The purpose of the so added instructions is only to create additional constraints. To not alter the semantics such as state or control flow of the actual contract the execution of inserted code is isolated from the rest. When the symbolic execution reaches a state that violates any constraint derived from an annotation, the contract is considered to violate the developer's expectations, and the violation is reported to the developer. \emph{Annotary} implements three types of annotations:\newline
\begin{enumerate}
  \item \textbf{Inline checks:} The annotation \lstinline[stringstyle=\color{darkgreen}]{"@check("BoolExpr")"} and its negation \\\lstinline[stringstyle=\color{darkgreen}]{"@never("BoolExpr")"} specify properties inside a contract function that are checks whether or not the specified condition holds. The condition can hold any boolean expression valid in solidity including calls to other functions and contracts.
  \item \textbf{Contract invariants:} The annotation \lstinline[stringstyle=\color{darkgreen}]{"@invariant("BoolExpr")"} defines a contract wide condition that has to hold whenever a transaction persists its state.
  \raggedright
  \item \textbf{Set restrictions:} restricts writing to a member variable from outside of explicitly allowed functions. A state at a \texttt{SSTORE} instruction is reported a violation if the \texttt{SSTORE} writes to the protected variable and the function is not explicitly allowed to. Users can specify these restrictions with the following annotation:\\
  \raggedright
  \lstinline[stringstyle=\color{darkgreen}]!"@set_restricted("["var="{[ContractName "."] MemberName[","]} ";"]} ["func="]{"constructor"|FunctionName|FunctionSignature}")"!
\end{enumerate}

\subsection{Modeling Transaction Execution}
Depending on the type of data, \emph{Annotary} uses different strategies to treat memory locations either as concrete or as symbolic values. Concrete values will be initialized according to the EVM's actual behavior, i.e., storage on contract creation will be initialized by \texttt{0}. Symbolic values refer to variables of the SMT constraint system which refer to specific memory locations. For instance, we write $\sigma[I_a]_s[key]\rightarrow BitVecRef(storage[key], 256)$ to denote the allocation of a variable \emph{key} in storage. In general, when writing data to some memory location, \emph{Annotary} supports both concrete and symbolic values and propagates data of the respective type to the memory location.
When data is read from a previously unused location, however, it depends on the data type whether \emph{Annotary} will treat it as symbolic or concrete:

\textbf{Call data} is modeled symbolically to represent all possible user interactions with the contract. \textbf{Memory} is treated concretely and reinitialized with the default value 0 for constructor and transaction execution. The \textbf{creation code} itself is known, but the appended initialization parameters are unknown at analysis time and thus handled symbolically. Reads after the end of the known instructions default to return symbolic variables.\newline
\textbf{Storage} is set to the concrete type when the constructor is executed. On this first transaction, the content of storage is known and defaults to returning 0 when reading from unwritten locations.

Then, storage is reset to be empty and treated symbolically henceforth to represent the most generic state space and account for all unknown transactions that might have happened between construction and invocation of the smart contract.

\subsection{Inter-contract Analysis}

\emph{Annotary} can correctly handle dependencies between contracts, including those which manifest only at Solidity but not at bytecode level.\newline
\textbf{Contract inheritance:} is the only relation that is not directly visible in bytecode and requires \emph{Annotary} to pre- and post-processes Solidity code. It uses the C3-linearization of the inheritance hierarchy to identify transaction implementations defined by a parent contract that are callable once the child contract is deployed.
Asserts referencing member variables of a child contract cannot be directly injected into the transaction function of the parent contract, as the member variable is not in the parent's scope.
To solve this, \emph{Annotary} generates a proxy function with the same signature in the child contract that delegates the call through the super keyword and injects the assert.\newline
\textbf{Nested contract creation:} Contracts can create other contracts by piggybacking the necessary constructor bytecode in their runtime bytecode. \emph{Annotary} spawns a new symbolic execution with the creation bytecode extracted from the current transaction execution.\newline
\textbf{Inter-contract interactions:} happen when the analyzed contract performs a message call or executes foreign contract code on their storage. Symbolically executing these interactions allows to resolve potentially returned values and to understand changes on the analyzed contracts state.
\emph{Annotary} implements symbolic execution for several EVM instructions for inter-contract interaction, lacking by Mythril, including \texttt{CREATE}, \texttt{STATICCALL}, \texttt{RETURNDATACOPY}, \texttt{RETURNDATASIZE}, and \texttt{EXTCODECOPY}. All instructions that trigger inter-contract interactions are executed with the appropriate concrete or symbolic persisting data type:\newline  
\textbf{\texttt{CALLCODE} and \texttt{DELEGATECALL}} execute external contract code referenced by the address of the external contract in the context of the current contract and can, therefore, change the contracts persistent storage. If address and code can be resolved, symbolic execution can account for these calls effects. If they cannot be resolved, storage has to be reset to be empty and symbolic, and the variables in the constraints are renamed to avoid collisions.\newline
\textbf{\texttt{CALL} and \texttt{STATICCALL}} are executed with empty symbolic storage for the first interaction and with the initialized symbolic storage on further interactions.\newline
\textbf{\texttt{CREATE}} deploys a new contract and executes the contract creation with empty concrete storage that is used for further interactions with the contract in the same transaction. In other transactions executions storage will be considered empty and symbolic.\newline
\subsection{Inter-transactional Analysis}
 \emph{Annotary} implements inter-transactional reachability analysis to eliminate false positive violations that are not reachable considering the possible set of contract transactions.

\subsubsection{Extracting Transaction Traces}\label{subsubsec:transactiontraces}
\emph{Annotary} uses transaction traces \textbf{$\tau = \{\Delta,\Phi\}$} for inter-transactional analysis, information of a contract execution that persists after the execution is finished:
\begin{itemize}
  \item $\mathbf{\Delta}$ is a mapping of symbolic state variables $k$ in $\sigma[I_a]$, e.g., storage slots or balance, of the currently analyzed contract to SMT bit vector expressions $\delta$, representing the change that a transaction performs on the state.
  \item The trace constraints $\mathbf{\Phi}$ are a set of conditions that have to hold such that the transaction represented by $\tau$ can be executed on the contract. $\mathbf{\Phi}$ is a subset of the path constraints. Path constraints with no reference to the previous state, e.g., only to input data, are not included in $\mathbf{\Phi}$ and do not lower the accuracy of reachability analysis.
\end{itemize}
Traces should represent state changing transactions that can appear amid a transaction sequence. The \texttt{global states} at the persisting instructions \texttt{STOP} and \texttt{RETURN}, are taken into consideration, while states at \texttt{SELFDESTRUCT} cannot be followed by further transactions and are therefore ignored. States with unchanged persisted values, e.g., in storage and balance, are filtered out due to irrelevance for the sequence and states with unsatisfiable path constraints due to inapplicability. \textbf{$\Delta$} and \textbf{$\Phi$} are extracted from global states that represent the unmodified contract execution, reducing constraints by all that are not relevant in an inter-transactional analysis.\newline
\emph{Annotary} differentiates between constructor transaction traces ($\bm{\tau_c}$) and message transaction traces ($\bm{\tau_m}$) and brings states that may violate an annotation into a transaction trace representation ($\bm{\tau_v}$) to allow reachability analysis.
\subsubsection{Chain Transaction Traces}\label{subsubsec:tracechaining}
\emph{Annotary} combines traces through expression substitution to explore the possible persisted states of a contract instead of iterative concolic execution. $\tau_{12} := \tau_1 \circ \tau_2$ represents the symbolic trace left onto the contract state when $\tau_1$ is executed before $\tau_2$. Definition~\ref{eq:tracechaining} shows how traces are combined.
The changes to the contract state $\Delta_1$ that trace $\tau_1$ applied exist at the beginning of trace $\tau_2$. Therefore the changes $\Delta_1$ have to be applied to the expressions used in $\Delta_2$ and $\Phi_2$.
\begin{align*} \label{eq:basetrace}
\tau_1 := \{\Delta_1,\Phi_1\}~~~~~
\tau_2 := \{\Delta_2,\Phi_2\}
\end{align*}
\begin{equation}\label{eq:tracechaining}
\tau_{12}~:=~\{\Delta_{12}:=\Delta_1 \circ_\Delta \Delta_2,~\Phi_{12}:=\Phi_1 \cup (\Delta_1 \circ_\Phi \Phi_2) \}
\end{equation}
$\circ_\Phi$ in Definition~\ref{eq:toConstraint} and $\circ_\Delta$ in Definition~\ref{eq:toStorage} are necessary operations to apply the storage changes $\Delta_1$ to $\tau_2$.
$\Phi$ is a list and $\Delta$ is a mapping of expressions. The pairs $(k',\delta')$ in the mapping $\Delta$ can be used together with the SMT-solvers \texttt{substitute} function to replace appearance of a value $k'$ in an expression e with $\delta'$.
\begin{equation}\label{eq:toConstraint}
\Delta\circ_\Phi\Phi~:=[substitute(\Delta,~\phi) : \phi \in \Phi]
\end{equation}
\begin{equation}\label{eq:toStorage}
\Delta_1\circ_\Delta\Delta_2~:=[(k,~substitute(\Delta_1,~\delta)):(k,~\delta) \in \Delta_2]
\end{equation}
We further define two properties for traces, spanning one or more transactions:
A trace $\tau$ is \textbf{valid} if its constraints are satisfiable. An invalid trace means that the constraints are not satisfiable and thus this sequence of instructions and calls among transactions is not executable at runtime. In the following, we denote satisfiability of a trace as $\textsf{sat}(\tau)$. A trace can further be \textbf{state independent} if its constraints do not contain any symbolic variables k referencing the previous contract state. State independence means that execution of that trace does not depend on the prior execution of any other contracts and is denoted by $\textsf{svar}(\tau) = \emptyset$.

\subsubsection{Confidence Levels}\label{subsubsec:severitylevel}
By combining the properties of validity and state independence, \emph{Annotary} expresses the confidence with which found violations will exist at runtime. \emph{Annotary} supports the following confidence levels, from most to least confident:
\begin{enumerate}
  \item \textbf{Single transaction violation}: For the intra-transactionally verified violating trace $\tau_v$, $\textsf{sat}(\tau_v)\land \textsf{svar}(\tau_v) = \emptyset$ holds. In this case, the transaction violates the annotation.
  \item \textbf{Chained transaction violation}: For a valid and state independent trace $\tau_{m^*v}$ of optionaly many applications\footnote{shorthand notation: $\tau_a \circ \tau_b:=\tau_{ab}$, application of $d$ arbitrary set members $\tau_{m^d}:=\tau_{m_1} \circ ... \circ \tau_{m_d}$} of transactions from $\tau_m$ and finally $\tau_v$ (i.e., $\textsf{sat}(\tau_{m^*v})\land \textsf{svar}(\tau_{m^*v}) = \emptyset$ holds). In this case, the annotation is violated, independent from which contract state the call is made.
  \item \textbf{Constructed violation}: A sequence of traces starting from the constructor was found that can trigger the annotation violation and is $\textsf{sat}(\tau_{cm^{*}v})\land \textsf{svar}(\tau_{cm^{*}v}) = \emptyset$. The attacker requires to be the contract creator or find a contract in the required state.
  \item \textbf{Unconfirmed violation}: The chaining depth $d$ was reached and there is at least one $\tau_{m^{d}v}$ that is $\textsf{sat}(\tau_{m^{d}v})\land \textsf{svar}(\tau_{m^{d}v}) \neq \emptyset$.
  \item \textbf{Violation avoiding context}: A point in the analysis was reached where the possibilities of chaining transactions to reach the violating state was exhausted. This means that although $\textsf{sat}(\tau_v)\land \textsf{svar}(\tau_v) \neq \emptyset$ it is also such that $\exists c \in \mathbb{N}: c <= d: (\nexists\tau_{m^{c}v}\in T:\textsf{sat}(\tau_{m^{c}v}))\land\forall e \in \mathbb{N}: e <=c:\nexists\tau_{m^{e}v}\in T:\textsf{sat}(\tau_{m^{e}v}) \land \textsf{svar}(m^{e}v)=\emptyset$.
  \item \textbf{Unsatisfiable violation}: The violating transaction $\tau_v$ is not satisfiable. This means $!\textsf{sat}(execution\_constraints_{\tau_v})$.
\end{enumerate}
\subsubsection{Chaining Strategy}\label{subsubsec:chainingstrat}
To check the validity of a found violation two high level strategies can be used to explore transaction traces:\newline
\textbf{Forward:} Starting from the set of constructor and transaction traces $T_c \cup T_m$, the current trace chains are applied to $T_v \cup T_m$, and the new trace chain is checked for satisfiability. If the chain is satisfiable, the violation is confirmed.\newline
\textbf{Backward:} Starting from the violating traces $\tau_v \in T_v$, the set of contract traces $T_c \cup T_m$ are applied to the set of remaining transaction chains. If an explored trace chain is valid and state independent, the violation is confirmed. If the form of the sequence is $\tau_{cm^*v}$ the exploration attempts to find a more threatening sequence $\tau_{m^*v}$.
\begin{figure}[!h]
\centering
\begin{tikzpicture}[node distance=0.5em and 0.1em]
  \node (tv1) {$\mathbf{\tau_{v1}}$};
  \node[right= of tv1, xshift=20em] (tv2) {$\mathbf{\tau_{v2}}$};

  \node[above left=of tv1] (tv1d) {...};
  \node[above right=of tv1] (tv1m2) {$\tau_{m2}\circ\mathbf{\tau_{v1}}$};

  \node[above left=of tv2] (tv2m1) {$\tau_{m1}\circ\mathbf{\tau_{v2}}$};
  \node[above right=of tv2] (tc1d) {...};

  \node[above left=of tv1m2] (tv1m2m1) {$\underline{\tau_{m1}\circ\tau_{m2}\circ\mathbf{\tau_{v1}}}$};
  \node[above right=of tv1m2] (tv1m2d) {...};

  \node[above left=of tv2m1] (tv2m1c1) {$\underline{\tau_{c1}\circ\tau_{m1}\circ\mathbf{\tau_{v2}}}$};
  \node[above right=of tv2m1] (tv2m1d) {...};

  \draw[thick, ->] (tv1) -- (tv1d);
  \draw[thick, ->] (tv1) -- (tv1m2);
  \draw[thick, ->] (tv2) -- (tv2m1);
  \draw[thick, ->] (tv2) -- (tc1d);
  \draw[thick, ->] (tv1m2) -- (tv1m2m1);
  \draw[thick, ->] (tv1m2) -- (tv1m2d);
  \draw[thick, ->] (tv2m1) -- (tv2m1c1);
  \draw[thick, ->] (tv2m1) -- (tv2m1d);

\end{tikzpicture}
\caption{Backward strategy finding state independent sequence for both violations.}\label{fig:backwardchaining}
\end{figure}
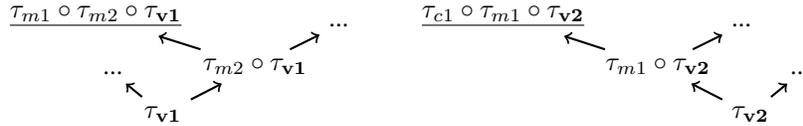
\emph{Annotary} uses this strategy depicted in Figure~\ref{fig:backwardchaining} as it allows to differentiate between violations with confidence level \textbf{unconfirmed violation} and \textbf{violation avoiding context}. The size of initial traces is smaller if $|T_c \cup T_m| > |Tv|$. Trace chaining scales better if $|T_c \cup T_m| < |T_v \cup T_m| \iff |T_c| < |T_v|$.

\section{Implementing Annotary}
This section elaborates on the implementation of the \emph{Annotary}, i.e., a Sublime Text plugin and the inter-contract concolic analysis on top Mythrils Laser-SVM.

\subsection{Preprocessing of Solidity Contracts}\label{subsec:parseData}
In a first preprocessing step, \emph{Annotary} parses Solidity source files and extracts the annotations stating the conditions that will be analyzed.
The input files are then parsed with the solidity compiler \texttt{solc} to gain the construction (bin) and runtime binaries (bin-runtime) of the contracts, the source code mappings (srcmap), that link the symbolically executed instructions to the code segments they were compiled from, the contracts application binary interface (ABI), which describes the transactions that can be executed and the expected input parameters, and finally the contract's abstract syntax tree (AST) to identify transaction endpoints, retrieve inheritance structures, functions and member variables.

\emph{Annotary} adds a \emph{rewriting} pass to the compilation process that converts \texttt{@check} and \texttt{@invariant} annotations into corresponding sets of \texttt{assert} statements, as in Appendix~\ref{sec:coderewritings}.
Furthermore, \emph{Annotary} modifies the \textsc{Laser}-SVM to isolate the execution of rewritten code from affecting the rest of the symbolic execution. We extended the \textsc{Laser}-SVM by a state processor that keeps track of instructions, result of the rewrite pass, excluding the resulting states from the set of unmodified contract states.

\subsection{Concolic Execution}\label{subsec:lasersvm}
\emph{Annotary} builds upon the \textsc{Laser}-SVM to extend inter-contract and adds inter-transactional analysis. We now guide the reader through the most significant building blocks that \emph{Annotary} adds to \textsc{Laser}-SVM.
\subsubsection{Symbolic Handling of Inter-contract Calls}\label{subsubsec:} \emph{Annotary} extends \textsc{Laser}-SVM by adding handlers for instructions that were not supported, in order to close the semantic gap between \textsc{Laser}-SVM and EVM. The \textbf{\texttt{CREATE}} instruction is implemented by executing a contract creation transaction with the nested contract code extracted from the current contract. The prior transaction execution is resumed with the newly created contract in $\sigma$. Support for the \textbf{\texttt{STATICCALL}} instruction is added, analog to the \texttt{CALL} instruction with \texttt{msg.value} set to 0, as no funds are transferred, and a flag that prevents \texttt{SSTORE}-instruction in nested calls to write persistent storage. \textbf{\texttt{RETURNDATASIZE}} and \textbf{\texttt{RETURNDATACOPY}} are implemented by extracting size and data from the global state and copying them to stack and memory, respectively. If \emph{Annotary} successfully resolves the concrete address that is given to \textbf{\texttt{EXTCODECOPY}}, the retrieved external code is copied to memory and treated concretely.

\subsubsection{Pre-, and Post-processing and Filtering of States}\label{subsubsec:stateprocessor}
\emph{Annotary} modifies how the \textsc{Laser}-SVM processes instructions in its worklist to handle instructions differently that have been added to the unmodified contract code.

For instance, the \texttt{ASSERT\_FAIL} instruction, would immediately terminate the execution when a given condition is not fulfilled. If that instruction has been inserted into the bytecode as a consequence of an \texttt{@check} annotation, however, we need a different semantic, as we want to detect the violation of the annotation, but not necessarily terminate the symbolic execution at that state. We thus extend \textsc{Laser}-SVM by \emph{state labels} that mark individual states in the explored symbolic state space as \emph{Violating} and/or \emph{Ignore} to indicate that this state was the result of code modification to identify violations and shall be saved and/or isolated from the set of states representing the execution of the unmodified code.

\subsection{Violation Identification and Classification}\label{subsec:outputlayer}
After the concolic execution, \emph{Annotary} has access to the state space of the unmodified contract and a set of states violating the \texttt{@check} or \texttt{@invariant} annotations. We can map these directly to warnings that will be displayed to the developer at the corresponding line of code. The \texttt{@set\_restricted} annotation limits write operations to a member variable to a set of valid functions and thus needs to search for violating states at \texttt{SSTORE} instructions.
The writing function is identified over the instruction association to code and the saved path constraint that stems from the selected function identifier. The written Solidity member is identified through the storage index according to the computed outline in storage, which requires \emph{Annotary} to keep track of relevant \texttt{keccak256} results, depicted in Appendix~\ref{subsec:algorithms}.
\emph{Annotary} searches through the state space for a path that starts at a violating location and ends in a state at a \texttt{STOP} or \texttt{RETURN} that would effectively persist the violating transaction to storage.

\subsubsection{Chaining Transactions}\label{subsubsec:buildtraces}
\emph{Annotary} chains transactions by merging selected states from the symbolic state space of one transaction into the state space of the previous transaction. The selection includes only relevant states, i.e. only those which have inter-transactional effects (e.g., write to storage).

When creating execution traces of transaction sequences, \emph{Annotary} maintains meta-data that is assigned to the sequence and presents users more qualified information such as the \textit{transaction depth} and the sequence of \textit{contract functions}, as well as data to optimize the trace chaining operation, such as the set of \textit{symbolic state variables} that references prior contract states and the set of \textit{transaction variables}. Both sets allow to keep track of variables that have to be substituted or renamed when combining transactions and allows for an efficient implementation using the Z3 expression substitution functionality. However, chaining transactions will lead to an explosion of possible chains and thus explored states, limiting the depth of transaction sequences that can be analyzed. We illustrate this effect with real-world contracts in \autoref{sec:evaluation}.\\
\emph{Annotary} analyzes transactions preceeding a violating trace for two purposes. A valid chain confirms the inter-transactional validity of the violation and the resulting confidence level gives a more nuanced judgment of the violation. Algorithm~\ref{lst:chainingstrat} in \autoref{subsec:algorithms} shows how the sequence of preceeding transactions with the highest confidence level that leads to a given violation is found. When the violating transaction sequence contains no symbolic state variables in the constraint expressions, a state independent chain of length one is found, it is assigned the confidence level \textit{single transaction violation}.
At every iteration step, all traces in $\tau_c \cup \tau_m$ are applied to the violating sequences of depth $n - 1$, which in the beginning is only the violating trace. Traces are only applied to a sequence if they overwrite a symbolic state variable and the set of constraints are checked for satisfiability before they are added to the set of violating sequences of length $n$.
If a transaction sequence ends in a constructor trace $\tau_c$, the chain is saved with the confidence level \textit{Constructed violation} but the search for a more severe violating sequence is continued. If the trace is from the set $\tau_m$ and chained trace is state independent, the search is terminated with a sequence of confidence \textit{Chained transaction violation}. If the set of new sequences is empty because all trace applications resulted in sequences with unsatisfiable constraints, the initial violation gets the confidence level \textit{Violation avoiding context}. If the maximal depth is reached the violating trace is of confidence level \textit{Unconfirmed violation}. After all violations are categorized, the annotation's violation confidence level is set to the highest level of the found sequences. Finally, all annotations with their violations are returned in JSON-format to the Sublime Text plugin.

\subsection{Annotary Plugin}\label{subsec:annotaryPlugin}
The \emph{Annotary} plugin bridges the gap to concolic execution from within the Sublime Text editor. Annotations are written inside of Solidity files, and a context menu allows to run the search for violations. Annotations and violating code pieces are visualized inside of the documents, e.g., in Figure~\ref{fig:annotation} in \autoref{sec:ideplugin}. Hovering over them shows the confidence level, the violating transactions, and informative description. A config allows disabling trace chaining, set the depth of chained traces and followed jumps during concolic execution.

\section{Discussion}\label{sec:evaluation}
We evaluated \emph{Annotary} with respect to its effectiveness and efficiency. First, we assessed how \emph{Annotary} performs with vulnerable contracts and if it would detect all vulnerabilities as expected. We thus created a sample set of 11 \textbf{small} contracts with known programming mistakes that have led to severe vulnerabilities in the past and added annotations that would have made \emph{Annotary} detect the flaw. Among others, this set includes the following mistakes: one of the Parity bugs \cite{paritymultisig2} allowed execution of an initialization function because of the unset member variable \texttt{initialized}. By adding an \texttt{@invariant(initialized==true)} annotation, \emph{Annotary} was able to spot this vulnerability. This mistake is especially hard to spot for humans if non-obvious call paths over library functions allow the execution of the initialization function~\cite{paritymultisig1} or if typos such as \texttt{state =+ 1} (which evaluates to \texttt{state = 1}) instead of \texttt{state += 1} are present. A further included mistake is to erroneously expose functions that allow writing to some member variable by incorrectly setting (or omitting) one of Solidity's four visibility modifiers for functions. \emph{Annotary} catches this error, if the member variable is annotated with \texttt{@set\_restricted}. Another, especially subtle mistake is writing to uninitialized structs. Structs can be persisted in either memory or storage and if declared in ''C style'' and not marked otherwise, default to storage. If fields of a struct are written without prior initialization, the write operation will overwrite the first storage slots, which can lead to disastrous consequences. Consider this snippet, which overwrites the owner address by calling \texttt{doSth()}.
\begin{lstlisting}[basicstyle=\scriptsize]
contract test{
  struct MyStruct { uint myField; }
  address owner;  // Keeps track of privileged owner

  function doSth() {
    MyStruct s;
    s.myField = uint(msg.sender); // Overwrites owner
  } }
\end{lstlisting}
\emph{Annotary} detects this vulnerability, if \emph{owner} is annotated with \texttt{@set\_restricted} (cf. \autoref{fig:annotation}), even taking delegated calls into account.\emph{Annotary} detects all programming mistakes in the ''small' sample set. Table~\ref{tab:mistakes} lists the mistakes and used annotation types.

In a second step, we were interested in the performance of \emph{Annotary} with real-world contracts and created a second sample set of 24 \textbf{large} contracts with the highest balance of Ether and available source code in the public Ethereum network. These contracts were not annotated and are not known to contain vulnerabilities, it is therefore not possible to create data underpinning the soundness and completeness of \emph{Annotary}. Nevertheless, we evaluated the coverage of the symbolic execution, the runtime, and scalability with respect to the depth of chained execution traces to give an impression on \emph{Annotary's} runtime. As can be seen from \autoref{tab:runtimecontracts}, the coverage of the ''large'' sample set is 80\%, while the coverage of the \textbf{small} set is 88\%. The average runtime for the ''small'' set is 4 seconds, which we consider well-suited for IDE integration, especially when considering that no performance optimizations have been done so far. For the real-world contracts from the ''large'' sets, the average runtime is with 700 seconds significantly higher due to larger code sizes. Columns \textbf{d$<$n$>$} in \autoref{tab:runtimecontracts} illustrate how increasing the depth of the analyzed call chains adds significant runtime overhead. A feasible mode of operation might thus be to configure the depth of analysis in the IDE to be lower and to run a full analysis in a CI server.

\begin{table}[!h]
  \caption[Evaluation of contracts.]{Average runtime of \emph{Annotary's} analysis of the ''small'' and ''large'' sample}
    \begin{adjustbox}{max width=\textwidth}
    \begin{tabular}{|c|c|c|c|c|c|c|c|c|c|c|}
    \hline
    \textbf{Type} &\textbf{Sample size} & \textbf{Coverage[\%]} &  \textbf{Sym. Exe.[s]} & \textbf{d1[s]} & \textbf{d2[s]} & \textbf{d3[s]}& \textbf{d4[s]} & \textbf{d5[s]} & \textbf{d6[s]} \\ \hline
    Small & 11 & 88 & 1.3  & 0.07 & 0.13 & 0.34 & 0.82 & 1.9 & 4.1\\\hline
    Large & 24 & 80 & 54.2 & 12.9 & 17.8 & 606  & - & - & - \\\hline

    \end{tabular}
  \end{adjustbox}
    \label{tab:runtimecontracts}
\end{table}

\section{Related Work}
Symbolic execution approaches with the pioneer Oyente~\cite{oyente} by Luu et al., the extension Osiris~\cite{osiris} by Torres et al., and Mythril~\cite{mythril} by Bernhard Mueller et al. use the results of symbolic execution and SMT-solving to find known vulnerabilities in an intra-transactional context. MAIAN~\cite{maian} by Nikolic et al. extends this approach to an inter-transactional context and finds vulnerability patterns defined over multiple transactions. \emph{Annotary} builds upon Mythril's \textsc{Laser}-SVM and extends it to support inter-contract analysis. Our work further differs from the aforementioned tools in that it supports inter-transactional executions chains. To the best of our knowledge, Teether~\cite{teether} by Krupp and Rossow is the only publication that also considers transaction traces. However, Teether does not allow customizable checking of properties but rather searches already deployed contract for a single vulnerability pattern. Our contribution is thus the first customizable development framework for smart contract developers, supporting inter-contract and inter-transactional analyses.
Further work related to our is Zeus~\cite{Kalra2018} by Kalra et al., which translates Solidity into LLVM and performs model checking against policies. Vandal~\cite{vandal} by Brent et al. converts EVM bytecode to abstract semantic logic relations and analyzes logic constraints over them. Formal verification in the form of Why3 \cite{why3} was already integrated into the Solidity online IDE Remix, requiring developers to create semi-assisted proofs, but the support was later removed~\cite{why3removed}. Other attempts include the formalization of contracts and EVM in F*~\cite{Bhargavan2016} by Bargavan et al. and in the formal verification framework Lem~\cite{lem}, that do not precisely capture inter-contract analysis and do not support inter-transactional analysis. Ahrendt et al. propose to translate Solidity into Java to make use of KeY, a well-approved theorem proving framework for Java programs \cite{Ahrendt2019}. Hildenbrandt et al. introduced the KEVM~\cite{kevm}, an executable formal specification of the EVM in the $\mathbb{K}$ framework which provides inter-contract and inter-transactional provability of claims by formulating all-path reachability statements. All these approaches require users to formulate desired properties in a formal language understood by the verifier, e.g., $\mathbb{K}$'s XML-style language or Why3's WhyML.

\section{Conclusions}
The field of secure development of smart contracts is still in its infancy and some of its challenges are fundamentally different from traditional software development due to the distributed computation model and the immutability of code. We contribute \emph{Annotary} to this field, an approach that strikes a balance between rigid but hard-to-use formal methods and static source code analyzers which have no knowledge of intent, thus producing too many false positives.
Our three main conclusions from this work are that first, annotations are a feasible way for developers to express their expectations and check their contracts for correctness in a language and environment they are comfortable with. Earlier work on integrating formal verification methods into Solidity has been dismissed for that reason, while the SMT-checking based approach that also \emph{Annotary} adopts seems to be well received by the community (cf. \cite{mythril,EthereumSMT2019}). Second, inter-contract and inter-transactional analysis are required to make sound statements about the security of a contract. Analysis of a single contract captures only a fraction of an actual Ethereum transaction and will not be able to create sound statements about safety and security guarantees. Third, the use of concrete values helps to increase precision and at the same time limit the complexity of the analysis. In contrast to traditional programs, where the specific execution environment is not known at the time of analysis, we can resolve concrete addresses referring to the Ethereum network and retrieve actual values from there.

The different confidence levels of \emph{Annotary} allow for a more nuanced interpretation of findings by the developer, as opposed to traditional source code analyzers which rank all findings equally relevant.
As part of our prototype evaluation, we have shown how \emph{Annotary} detects common programming pitfalls and is able to detect cross-transaction vulnerabilities. The runtime analysis suggests its applicability in an IDE for smaller contracts and acceptable runtimes for larger contracts when integrated into continuous integration (CI) processes.

\subsubsection*{Acknowledgements}
This work was partially funded by the Bavarian Ministry of Economics as part of the initiative Bayern Digital as well as the Fraunhofer Cluster of Excellence "Cognitive Internet Technologies".

%
%
%
\bibliographystyle{splncs04}
\bibliography{preprint}

\newpage
\appendix
\section{Appendix}\label{sec:appendix}

\subsection{Annotary IDE Plugin}\label{sec:ideplugin}

\begin{figure}[h!]
  \resizebox{.8\textwidth}{!}{
  \includegraphics[width=\linewidth]{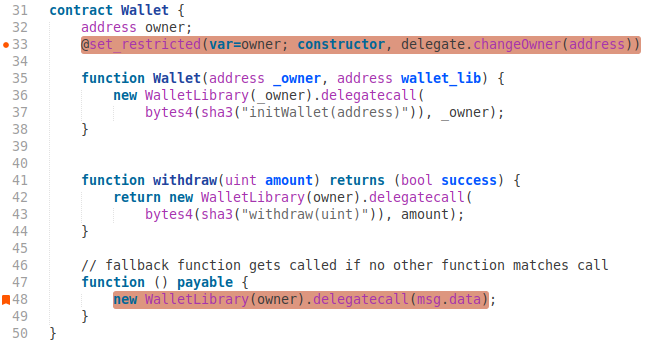}
  }
  \caption[Annotation information]{Annotary marks violated annotations and violating code.}
  \label{fig:annotation}
\end{figure}

\subsection{Code Rewritings}\label{sec:coderewritings}

\paragraph{Inline Checks}
\emph{at annotation position:} \texttt{@check(condition) $\longrightarrow$ assert(condition);}

\paragraph{Asserting Invariants}
  \emph{- at empty block end:} \texttt{$\emptyset \longrightarrow$ assert(condition);}.
  \newline\emph{- before empty return statement:} \texttt{return; $\longrightarrow$ assert(condition); return;}.
  \newline\emph{- before return with value:} \texttt{return (exp1, ...); $\longrightarrow$ var (v\_<nonce1>,...) = (exp1, ...; assert(condition); return (v\_<nonce1>, ...);}.

\paragraph{Proxy Asserts to inherited Functions that}

  \emph{- do not return values:} $\emptyset \longrightarrow$ \lstinline|function f_name(param1, ...)...| \lstinline| { super.f_name(param1, ...);  assert(condition);}|
  \newline\emph{- do return values:} $\emptyset \longrightarrow$ \lstinline|function f_name(param1, ...)...{|\newline \lstinline|var (v_<nonce1>,...) = super.f_name(param1, ...);|\newline
  \lstinline|assert(condition);  return (v_<nonce1>, ...); }|

\subsection{Algorithms}\label{subsec:algorithms}

\begin{lstlisting}[,caption={Code added to the \texttt{ADD}-instruction to keep track of expression involved in index and mapping key computations.},captionpos=b, label={lst:addmap}, style=base]
if o1 in keccakMap or o2 in keccakMap:
  keccakMap[simplify(o1 + o2)] = get(keccakMap, o1) + get(keccakMap, o2)
\end{lstlisting}
\begin{lstlisting}[,caption={Code added to the \texttt{SHA3}-instruction to keep track of expression involved in index and mapping key computations.},captionpos=b, label={lst:sha3}, style=base]
for word in input:
  if word in keccakMap:   word = keccakMap[word]
  if result in keccakMap:   keccakMap[result] = Concat(keccakMap[result], word)
  else:   keccakMap[result] = word
\end{lstlisting}

\begin{lstlisting}[,caption={Algorithm to determine the severity level in a violating trace by analyzing the inter-transaction reachability. } ,captionpos=b, label={lst:chainingstrat}, escapeinside={(*}{*)}]
check_severity(v, (*$T_c$*), (*$T_m$*), max_d, pref_ind):
  T, (*$\tau_{cv}$*) := (*$T_c \cup T_m$*), (*$\perp$*)
  if v.status == VSINGLE:
    return v, VSINGLE
  (*$VS$*) := Queue(v)
  for d in {1..max_d}:           (*$\triangleleft$*) run until max depth
    (*$VS_{new}$*) := Queue(v)
    while (*$VS$*) != (*$\emptyset$*):
      vs := (*$VS$*).pop()
      for (*$\tau \in T$*)
        if (*$\tau_{cv} != \perp \wedge \tau \in T_c$*):
          continue           (*$\triangleleft$*) skip construction traces
        if (*$\tau.storage.keys \cap v.storage\_vars$*) != (*$\emptyset$*):
          vt = (*$\tau \circ vs$*)   (*$\triangleleft$*) apply trace
          if (*$vt == \perp$*):
            continue            (*$\triangleleft$*) Chain not satisfiable
          if (*$\tau \in T_c$*):
            zeroize_storage_vars(vt)
            if not satisfiable(vt.constraints)
              continue  (*$\triangleleft$*) zeroize and check const. trace
          if sym_storage_vars(vt.constraint) == (*$\emptyset$*):
            if not pref_ind (*$\wedge~\tau \in T_c$*):
              (*$\tau_{cv}$*) := vt, VCHAIN (*$\triangleleft$*) save found const. trace
            else:
              return vt, VCHAIN (*$\triangleleft$*) found violating chain
          else:
            (*$VS_{new}$*).push(vt)  (*$\triangleleft$*) save open state
    if (*$VS_{new}$*) == (*$\emptyset$*):          (*$\triangleleft$*) trace chain space exhausted
      if (*$\tau_{cv}$*) != (*$\perp$*):
        return (*$\tau_{cv}$*)
      else:
        return (*$\perp$*), HOLDS
    else:
      (*$VS$*) := (*$VS_{new}$*)
  if (*$\tau_{cv}$*) != (*$\perp$*):              (*$\triangleleft$*) max depth reached
    return (*$\tau_{cv}$*)
  else:
    return (*$VS$*).pop() , VDEPTH



\end{lstlisting}

\begin{table}[!h]
  \centering
    \caption{Uncovered implementation mistakes in ''small'' sample with annotation types.}
    \begin{adjustbox}{max width=\textwidth}
    \begin{tabular}{|c|c|c|c|}
    \hline
    \textbf{Mistake} &\textbf{Uncovering annotation type} \\ \hline
    Over-/Underflow & @invariant \\\hline

    Struct cast to storage & @set\_restricted \\\hline

    Misspelled constructor name & @set\_restricted\\\hline

    Missing visibility modifier & @invariant \& @set\_restricted \\\hline

    Memory layout missmatch with delgation & @set\_restricted \& @check \\\hline

    Unmatched call forwarded to delegate  & @set\_restricted \\\hline

    Unset state (instanciated) & @invariant \\\hline
    Unchecked send return & @check \\\hline
    Arithmetic mistace (=+) & @check \\\hline
    Trick transaction origin & @invariant \\\hline
    Unreachable state/code & @invariant \\\hline

  \end{tabular}
  \end{adjustbox}
    \label{tab:mistakes}
\end{table}
\end{document}